\documentclass[aps,prb,twocolumn,superscriptaddress,showpacs,amsmath,amssymb,longbibliography]{revtex4-1}
\usepackage{graphicx}
\usepackage[breaklinks]{hyperref}

\usepackage[utf8]{inputenc}
\usepackage[english]{babel}
\usepackage{amsmath}
\usepackage{xcolor}

\definecolor{nred} {RGB}{224,0,0}
\definecolor{nblue} {RGB}{28,130,185}

\begin{document}

\title{Similarity of thermodynamic properties of Heisenberg model on
  triangular and kagome lattices}  
\author{P. Prelov\v{s}ek}
\affiliation{Jo\v zef Stefan Institute, SI-1000 Ljubljana, Slovenia}
\affiliation{Faculty of Mathematics and Physics, University of Ljubljana, SI-1000 
Ljubljana, Slovenia}
\author{J.  Kokalj}
\affiliation{Faculty of Civil and Geodetic Engineering, University of Ljubljana, SI-1000 Ljubljana, Slovenia}
\affiliation{Jo\v zef Stefan Institute, SI-1000 Ljubljana, Slovenia}

\begin{abstract}
Derivation of a reduced effective model allows for a unified
treatment and discussion of the $J_1$-$J_2$ $S=1/2$ Heisenberg model on 
triangular and kagome lattice. Calculating thermodynamic quantities,
i.e. the entropy $s(T)$ and uniform susceptibility $\chi_0(T)$,
numerically on systems up to effectively $N=42$ sites, we show by comparing to 
full-model results that low-$T$ properties are qualitatively well represented
within the effective model.  Moreover, we find in the spin-liquid
regime similar variation of $s(T)$ and $\chi_0(T)$ in both models
down to $T \ll J_1$. In particular, studied spin liquids appear to be
characterized by Wilson ratio vanishing at low $T$, indicating that 
the low-lying singlets are dominating over the triplet excitations.

\end{abstract}

\maketitle

\section {Introduction} 

Studies of possible quantum spin-liquid
(SL) state in spin models on frustrated lattices have a long history,
starting with the Anderson's conjecture \cite{anderson73} for the
Heisenberg model on a triangular lattice. In last two decades
theoretical efforts have been boosted by the discovery of several
classes of insulators with local magnetic moments
\cite{lee08,balents10,savary17}, which do not reveal long-range order
(LRO) down to lowest temperatures $T$.  The first class are compounds,
as the herbertsmithite ZnCu$_3$(OH)$_6$Cl$_2$ \cite{norman16}, which
can be represented with Heisenberg $S=1/2$ model on kagome lattice,
being the subject of numerous experimental studies
\cite{mendels07,olariu08,han12,fu15}, now including also related
materials \cite{hiroi01,fak12,li14,gomilsek16,feng17,zorko19}
confirming the SL properties, at least in a wide $T > 0$
range. Another class are organic compounds, as
$\kappa$-(ET)$_2$Cu$_2$(CN)$_3$
\cite{shimizu03,shimizu06,itou10,zhou17}, where the spins reside on a
triangular lattice.  Recently, the charge-density-wave system
1T-TaS$_2$, which is a Mott insulator without magnetic LRO and shows
spin fluctuations at $T > 0$
\cite{klanjsek17,kratochvilova17,law17,he18}, has been added into this
family.

Numerical \cite{bernu94,capriotti99,white07} and analytical
\cite{chernyshev09} studies of the nearest-neighbor (nn) quantum
$S=1/2$ Heisenberg model (HM) on a triangular lattice (TL) confirm a spiral
long-range order (LRO) with spins pointing in $120^\circ$ tilted
directions.  Introducing the next-nearest-neighbor (nnn) coupling
$J_2>0$ enables the SL ground state (g.s.)  in the part of the phase
diagram
\cite{kaneko14,watanabe14,zhu15,hu15,iqbal16, wietek17,gong17,ferrari19}. There 
is even more extensive literature on the HM on the kagome lattice (KL)
confirming the absence of g.s. LRO order \cite{mila98,
  budnik04,iqbal11,lauchli11,iqbal13}.  The prevailing conclusion of
numerical studies of the g.s. and lowest excited states is that HM on
KL has a finite spin triplet gap $\Delta_t$
\cite{yan08,lauchli11,depenbrock12, lauchli19} (with some evidence
pointing also to gapless SL \cite{iqbal13,liao17,chen18}), but much
smaller or vanishing singlet gap $\Delta_s \ll \Delta_t $
\cite{elstner94,mila98,sindzingre00,misguich07,
bernu15,lauchli19,schnack18}. On the other hand, extensions into the
$J_1$-$J_2$ model \cite{kolley15,liao17} with $J_2 > 0$ again leads
towards g.s. with magnetic LRO. Still, HM on both lattices in their
respective SL parameter regimes have been studied and considered
separately, not recognizing or stressing their similarity.

Our goal is to put extended $J_1$-$J_2$ HM on TL and
KL on a common ground, stressing the similarity of their (in particular thermodynamic) 
properties within their presumable SL regimes.
To this purpose we derive and employ a reduced effective model (EM),
which is based on keeping only the lowest four $S=1/2$ states in a
single triangle.  Such an EM has been previously introduced and analyzed
for the case of KL \cite{subrahmanyan95,mila98,budnik04,capponi04} and as the starting 
point for block perturbation approach also for the TL \cite{subrahmanyan95}, 
but so far has not been used to evaluate and capture $T>0$ properties.  Such 
an EM has an evident advantage of reduced number of states in an
exact-diagonalization (ED) study and hence allowing for somewhat
larger lattices (in our study up to $N=42$ sites). Still, this is not the most important 
message, since EM allows also an insight into the character of low-energy excitations,
being now separated into spin (triplet) and chirality (singlet) ones. The main
focus of this work is on the numerical evaluation of thermodynamic
quantities and their understanding, whereby we do not resort to perturbative limits 
(extreme breathing limit) of weakly coupled triangles \cite{repellin17,iqbal18} which 
apparently does  not represent fully the same SL physics.  We concentrate on the entropy density 
$s(T)$ and uniform  susceptibility $\chi_0(T)$ within the SL parameter regimes, approached before mostly
by high-$T$ expansion \cite{elstner93,elstner94,sindzingre00,misguich07,rigol07,rigol07a,
bernu15,bernu19} and only recently with numerical methods adequate for lower
$T \ll J_1$, both on TL \cite{prelovsek18} and KL \cite{chen18,schnack18}. However, 
the most universal property is the temperature-dependent Wilson ration $R(T)$, introduced and 
discussed further on.

\subsection{Generalised Wilson ratio} 

Our results in the following reveal that
in both lattices and within the SL regime $s(T)$ and $\chi_0(T)$ are very
similar in a broad range of $T$. In this respect very convenient quantity is
$T$-dependent generalized Wilson ratio $R(T)$, defined as
\begin{equation}
R=  4 \pi^2 T \chi_0 / (3 s), \label{rw}
\end{equation}
which is equivalent (assuming theoretical units $k_B= g \mu_B =1$) to the
standard $T=0$ Wilson ratio in the case of Fermi-liquid
behavior where $s= C_V=\gamma T$. Definition Eq.(\ref{rw}) is more 
convenient with respect to the standard one (with $C_V$ instead of $s$)  
since it has meaningful $T$ dependence due to monotonously increasing
$s(T)$, having also finite high-$T$ limit
$R_\infty = \pi^2/(3 \ln2) = 4.75$.  Moreover, it can differentiate
between quite distinct $T \to 0$ scenarios: 

\noindent a) in the case of magnetic LRO at
$T \to 0$ one expects in 2D (isotropic HM)
$\chi_0(T\to 0) \sim \chi_0^0 >0$ but $s \propto T^2$~
\cite{manousakis91}, so that $R_0 = R(T \to 0) \to \infty$, 

\noindent b) in a
gapless SL with large spinon Fermi surface one would expect
Fermi-liquid-like constant $R_0 \sim 1$ \cite{balents10,zhou17,law17}, 

\noindent c) $R_0 \ll 1$ or a decreasing $R(T \to 0) \to 0$ would indicate that
low-energy singlet excitations dominate  over the triplet ones
\cite{balents10,lauchli19}. 

In the following we find in the SL regime
numerical evidence for the last scenario, which within the EM we
attribute to low-lying chiral fluctuations being a hallmark of Heisenberg model
in the SL regime. It should be pointed out that the same property might be 
very general property of SL models and it remains to be clarified in relation with
experiments on SL materials.

In Sec.~II we derive and discuss the form of the EM model for the $J_1$-$J_2$ Heisenberg 
model on both TL and KL. In Sec.~III we present numerical methods employed to calculate
thermodynamic quantities in the EM as well as the in full models. 

\section{Reduced effective model}
 
\begin{figure}[h!]
\centering
\includegraphics[width=1.1\columnwidth]{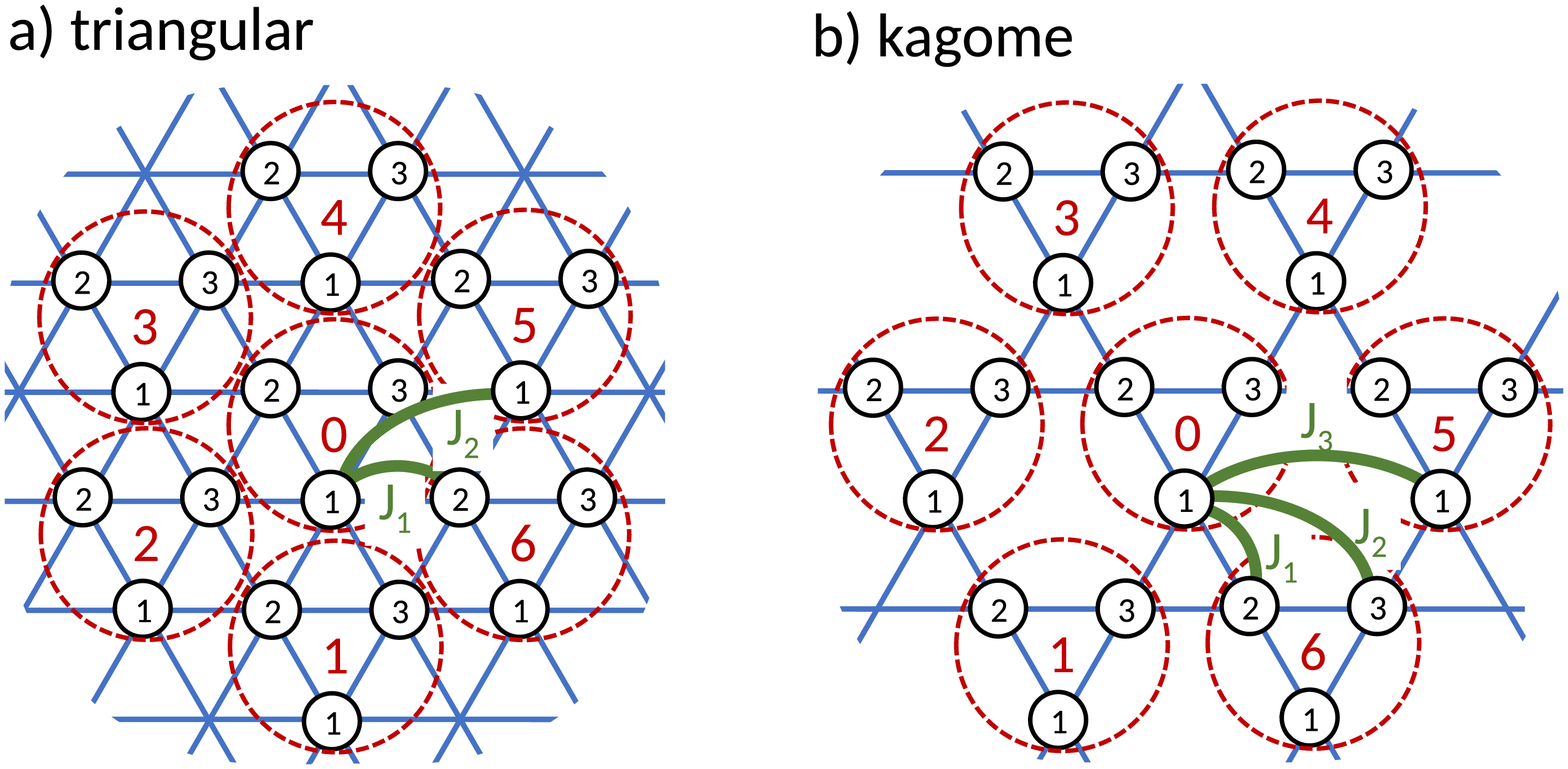}
\vspace{-1cm}
\caption{(a) Triangular and (B) kagome lattice represented as the (triangular) lattices
of coupled basic triangles.   Shown are also model exchange couplings on both lattices.}  \label{figsl0}
\end{figure}

We consider the isotropic $S=1/2$ extended $J_1$-$J_2$ Heisenberg
model,
\begin{equation}
H= J_1  \sum_{\langle kl \rangle} {\bf S}_k \cdot  {\bf S}_l + J_2  \sum_{\langle \langle kl \rangle \rangle} 
{\bf S}_k \cdot  {\bf S}_l, \label{hjj}
\end{equation}
on the TL and KL, where $J_1>0$ and $J_2 $ refer to nn and nnn
exchange couplings (see Fig.\ref{figsl0}), respectively.  The role of $J_2 >0$ on TL is to
destroy the 120$^0$ LRO allowing for a SL
\cite{kaneko14,watanabe14,zhu15,iqbal16,prelovsek18}, while for KL it
has the opposite effect \cite{kolley15}.  Further we set $J_1=J=1$ as an
energy scale. 

As shown in Fig.~\ref{figsl0} the model Eq.~(\ref{hjj}) on both lattices can be represented as
coupled basis triangles \cite{subrahmanyan95,mila98,budnik04,capponi04} where we keep
in the construction of the EM only four degenerate $S=1/2$ states (local
energy $\epsilon_0=-3/4$), neglecting higher $S=3/2$ states (local
$\epsilon_1=3/4$),
\begin{eqnarray}
|\uparrow \pm \rangle  &=&\frac{1}{\sqrt{3}}[ | \downarrow\uparrow \uparrow \rangle +
\mathrm{e}^{\pm i \phi} | \uparrow\downarrow \uparrow \rangle +\mathrm{e}^{\mp i \phi} | \uparrow\uparrow \downarrow \rangle ], 
\nonumber \\
|\downarrow \pm \rangle  &=& \frac{1}{\sqrt{3}}[ | \uparrow\downarrow \downarrow \rangle + \mathrm{e}^{\mp i \phi}   | \downarrow\uparrow \downarrow \rangle +
\mathrm{e}^{\mp i \phi} | \downarrow\downarrow \uparrow \rangle ], \label{updown}
\end{eqnarray}
where $\phi = 2 \pi/3$, $\uparrow, \downarrow$ are (new) spin states
and $\pm$ refer to local chirality.  One can rewrite Eq.(\ref{hjj})
into the new basis acting between nn triangles $\langle i,j \rangle$.
The derivation is straightforward taking care that the matrix elements of the original model.
Eq.~(\ref{hjj}), within the new basis, Eq.~(\ref{updown}), are exactly
reproduced with new 
operators.  We follow the procedure through intermediate (single triangle) orbital spin operators 
(see Fig.~\ref{figsl0}),
\begin{equation}
S_{(0,+,-)}=S_1+(1,\omega,\omega^*) S_2+ (1,\omega^*,\omega) S_3, \label{s0}
\end{equation} 
with $\omega =  \mathrm{e}^{ i \phi}$. Operators in Eq.~(\ref{s0}) have only few nonzero matrix elements within
the new basis, Eq.~(\ref{updown}), e.g., 
\begin{eqnarray}
&&\langle  \uparrow + | S_0^z | \uparrow + \rangle = -\langle  \downarrow + | S_0^z | \downarrow + \rangle = \frac{1}{2},  \nonumber \\
&&  \langle  \downarrow + | S_-^z | \downarrow - \rangle = - \langle  \uparrow + | S_- ^z | \downarrow + \rangle =  1 , \nonumber \\
&&\langle  \uparrow + | S_0^+ | \downarrow + \rangle = ~~\langle  \uparrow - | S_0^+ | \downarrow - \rangle = -1, \nonumber \\
&&\langle  \uparrow - | S_+^+ | \downarrow + \rangle = ~~\langle  \downarrow + | S_-^- | \uparrow - \rangle = 2. \label{locs}
\end{eqnarray}
Such operators can be fully represented in terms of standard local $s=1/2$ spin operators ${\bf s}$, 
and pseudospin (chirality) operators ${\bf \tau}$(again $\tau=1/2$),
\begin{eqnarray}
&&S^z_0= s^z,  \quad S^\pm_0=-s^\pm , \nonumber \\
&& S^z_\pm= -2 s^z \tau^\mp, \quad S^\pm_+=  2 s^\pm \tau^-, \quad S^\pm_-=  2 s^\pm  \tau^+,  \label{effspin}
\end{eqnarray} 
Since the original Hamiltonian, Eq.~(\ref{hjj}), acts only between two sites on the new 
reduced (triangular) lattice, the effective model (EM), subtracting local $\epsilon_0$,
can be fully represented for both lattices 
(as shown by an example further on) in terms of ${\bf s}_i$ 
and ${\bf \tau}_i$ operators, introduced in Eq.~(\ref{locs}),
\begin{eqnarray}
\tilde H &=& \frac{1}{2} \sum_{i ,d } {\bf s}_{i} \cdot  {\bf s}_{j} ( D+ {\cal H}^d_{ij})  , \label{em} \\ 
{\cal H}^d_{ij} &=& F_d \tau^+_i \tau^-_j + P_d \tau^+_i + Q_d \tau^+_j + T_d \tau^+_i \tau^+_j + \mathrm{H.c}, \nonumber
\end{eqnarray}
where directions $d=1$--$6$ and $j=i+d$ run over all nn sites of site $i$,
and the new lattice is again TL.  We note that Eq.~(\ref{em}) corresponds
to the one studied before for simplest KL \cite{subrahmanyan95,mila98,budnik04}, but it is
valid also for TL \cite{subrahmanyan95} and nnn $J_2$. It is remarkable that the spin part
remains $SU(2)$ invariant whereas the chirality part is not, i.e., it
is of the XY form. 

It should be pointed out that EM, Eq.~(\ref{em}), is not based on a perturbation expansion assuming weak 
coupling between triangles, although in the latter case it offers a full description of low-lying states and can be further 
treated analytically  in (rather artificial) strong breathing limit $|\tilde H| \ll |E_0|$ 
\cite{subrahmanyan95,mila98}. As for several other applications of reduced effective models (prominent 
example being the $t$-$J$ model as reduced/projected Hubbard model), one expects that low-$T$ physics is (at least
qualitatively) well captured by the EM. 

\subsection {Triangular lattice}

For the considered TL and KL we further present the actual parameters of the EM,
Eq.~(\ref{em}). The derivation is straightforward using the representation 
Eqs.~(\ref{s0}) and (\ref{effspin}). As an example we present the $J_1$-term interaction in
Eq.~(\ref{hjj}) between (new) sites $0$ and $1$ (see Fig.~\ref{figsl0}a),
\begin{eqnarray}
&&~~~~\tilde H^1_{01}/J_1 = {\bf S}_{01} \cdot ( {\bf  S}_{12} + {\bf  S}_{13} )  = \nonumber \\
&& = \frac{1}{9}(S_{00}^z + S^z_{0+} +  S^z_{0-})(2 S^z_{10} -  S^z_{1+} -  S^z_{1-} ) + \nonumber \\
&&+\frac{1}{18} [(S_{00}^+ + S^+_{0+} +  S^+_{0-})(2 S^-_{10} -  S^-_{1+} -  S^-_{1-})  + 
\mathrm{H.c.}]  = \nonumber \\
&&= - \frac{4}{9} {\bf s}_0 \cdot {\bf s}_1 (\tau^-_0 + \tau^+_0 -\frac{1}{2})(\tau^-_1 + \tau^+_1 + 1).
\label{h01}
\end{eqnarray}
Deriving in the same manner also other $\tilde H^d$ terms, including now also $J_2$ term 
(which are even simpler in TL), we can identify the  parameters (with $J_1=1$)  in the EM, Eq.(\ref{em}),
\begin{equation}
D=  \frac{2}{9} + \frac{1}{3} J_2, \qquad  F=  - \frac{4}{9} + \frac{4}{3} J_2, 
\label{df}
\end{equation}
and further terms depending explicitly on direction  $d$,
\begin{eqnarray}
P_1&=& -\frac{4}{9},\quad P_2= \frac{2}{9} \omega^* , 
\quad P_3= -\frac{4}{9} \omega , \nonumber \\
P_4&=&  \frac{2}{9} , \quad P_5= -\frac{4}{9} \omega^* ,
\quad P_6= \frac{2}{9} \omega, \nonumber \\
T_1&=& -\frac{4}{9} ,\quad T_2 = -\frac{4}{9} \omega , \quad T_3 = -\frac{4}{9} \omega^* 
\end{eqnarray}
with $T_{d+3}=T_d$ and $Q_d = P_{d+3}$. It is worth noticing, that $J_2$ does not enter
in terms $P_d,Q_d,T_d$.
It can be also directly verified, that for the latter couplings, the average over all nn bonds vanish, i.e.,
\begin{equation}
\bar P = \frac{1}{6}\sum_d P_d = 0, \qquad \bar Q = \bar T =0,
\end{equation}
indicating possible minor importance of these terms. This is, however, only partially true
since such terms also play a role of distributing the increase of entropy $s(T)$ to a wider $T$
interval. 

Eqs.~(\ref{em}),(\ref{df}) yield also some basic insight into the HM model
on TL, as well the similarity between models on TL and KL.  While $\chi_0(T)$ is
governed entirely by ${\bf s}$ operators, low-$T$ entropy $s(T)$ (and
specific heat $C_V(T)$) involves also chirality ${\bf \tau}$ fluctuations. In
TL at $J_2=0$ $\tau$ coupling is ferromagnetic and favors spiral
$120^0$ LRO. $\tau$ fluctuations are enhanced via $J_2 >0$ reducing
$F$ and finally $F \to 0$ on approaching $J_2 \sim 0.3$. Still, before
such large $J_2$ is reached $P_d,Q_d,T_d$ terms become relevant and 
stabilize SL at $J_2 \sim 0.1$ \cite{kaneko14,watanabe14,zhu15,hu15,iqbal16}.  It
should be stressed that in TL a standard magnetic LRO requires LRO ordering of
both ${\bf s}$ and ${\bf \tau}$ operators. 

\subsection{Kagome lattice.} 

In analogy to the TL example, Eq.~(\ref{h01}), we derive also the corresponding terms for the case of KL.
Without loosing the generality we can include here also the
third-neighbor exchange term $J_3$, 
see Fig.~\ref{figsl0}b,  which also couples only neighboring triangles.
Then $D$ and $F_d$ couplings are given by
\begin{equation} 
D =   \frac{1}{9}  + \frac{2}{9} J_2   + \frac{2}{9} J_3, \qquad  F_1 =   
\frac{4}{9} \omega  + \frac{8}{9} \omega^* J_2 + \frac{4}{9} J_3,
\end{equation}
while $F_{d+1}= F_d^*$. In contrast to TL, in KL at $J_2=0$ the $F_d$ coupling is complex and alternating, with a
nonzero  average, being real and negative, i.e. $\bar F = (1/6) \sum_d F_d < 0$.
Moreover, $| \mathrm{Im} F_d| < |\mathrm{Re} F_d |$, indicates the absence of
LRO.  Here, $J_2 >0$ reduces $|\mathrm{Im} F_d|$ and on approaching
$J_2 \sim 0.5$ one reaches real $F_d <0$ connecting KL model to TL at
$J_2=0$ and related LRO, as observed in numerical studies
\cite{kolley15}.

Further terms are given by
\begin{eqnarray}
P_1&=& - \frac{2}{9}  + \frac{2}{9} \omega^* J_2 - \frac{2}{9} \omega J_3, \quad 
P_2 = - \frac{2}{9} \omega  + \frac{2}{9} \omega^* J_2 - \frac{2}{9} J_3, \nonumber \\
P_3&=& - \frac{2}{9} \omega  + \frac{2}{9} J_2 - \frac{2}{9} \omega^* J_3, \quad
P_4 = - \frac{2}{9} \omega^*  + \frac{2}{9} J_2 - \frac{2}{9} \omega J_3, \nonumber \\
P_5&=& - \frac{2}{9} \omega^*  + \frac{2}{9} \omega J_2 - \frac{2}{9} J_3, \quad
P_6  = - \frac{2}{9}  + \frac{2}{9} \omega J_2 - \frac{2}{9} \omega^* J_3, \nonumber \\
\end{eqnarray}
with $Q_d=P_{d+3}$ and
\begin{eqnarray}
T_1&=&  \frac{4}{9} \omega^*  - \frac{4}{9} \omega^* J_2 + \frac{4}{9} J_3, 
\quad T_2 =  \frac{4}{9}   - \frac{4}{9}  J_2 + \frac{4}{9} J_3, \nonumber \\
T_3&=& \frac{4}{9} \omega  - \frac{4}{9} \omega J_2 + \frac{4}{9} J_3, \qquad T_{d+3}=T_d .
\end{eqnarray}
Again, terms which do not conserve $\tau^z_{tot}$ have the property $\bar P =\bar Q = \bar T =0$.

\section{Numerical method}

In the evaluation of thermodynamical quantities we use the FTLM
\cite{jaklic94,jaklic00,prelovsek13}, which is based on the Lanczos
exact-diagonalization (ED) method \cite{lanczos50}, whereby the Lanczos-basis states
are used to evaluate the normalized thermodynamic sum
\begin{equation}
Z(T) = \mathrm{Tr} \exp[- (H-E_0)/T],
\end{equation}
(where $E_0$ is the ground state energy of a system). The FTLM is
particularly convenient to apply for the calculation of the conserved
quantities , i.e., operators $A$ commuting with the Hamiltonian
$[H,A]=0$. In this way we evaluate $Z$, the thermal average energy
$\Delta E =\langle H - E_0 \rangle $ and magnetization
$M = \langle s^z_{tot} \rangle$. From these
quantities we evaluate the thermodynamic observables of interest,
i.e. uniform susceptibility $\chi_0(T)$ and entropy density $s(T)$,
\begin{equation}
\chi_0= \frac{M^2}{N T}, \qquad  s = \frac{T \ln Z +  \Delta E}{N T} ,
\end{equation}
where $N$ is the number of sites in the original lattice.

We note that for above conserved operators $A$ there is no need to store Lanczos
wavefunctions, so the requirements are essentially 
that of the g.s. Lanczos ED method, except that we need the summation over all
symmetry sectors and a modest sampling $N_s < 10$ over initial
wavefunctions is helpful. To reduce the Hilbert space of basis states $N_{st}$ 
we take into account  symmetries, in particular the translation symmetry 
(restricting subspaces to separate wavevectors $q$) and 
$s^z_{tot}$ while the EM, Eq.~(\ref{em}), does not conserve $\tau^z_{tot}$. 
In such a framework in the present study we are restricted to systems
with $N_{st} < 5 \cdot 10^6$ symmetry-reduced basis states, which means EM with up to
$N = 42$ sites. The same system size would require in the full HM 
$N_{st} \sim 10^{10}$ basis states.
  
An effective criterion for the macroscopic relevance of FTLM results is
$Z(T) \gg 1$ (at least for system where gapless excitations are
expected), which in practice leads to a criterion
$Z>Z^*=Z(T_{fs}) \gg 1$ determining the finite-size temperature $T_{fs}$.
Taking $Z^* \sim 20$ implies (for $N=42$)
also the threshold entropy density $s(T_{fs}) \sim 0.07$,
independent of the model. It is then evident that $T_{fs}$ depends crucially 
on the model, so that large $s(T)$ works in favor of
using FTLM for frustrated and SL systems. Here we do not present 
the finite-size analysis of FTLM results within EM, but they are quite analogous to previous 
application of the method to HM on KL \cite{schnack18} where similar low $T_{fs}$ was 
established (but much higher $T_{fs}$ ) in unfrustrated lattices and models   \cite{jaklic00}.
Moreover, in the case of models with a sizable gap, e.g. $\Delta > T_{fs}$ the results of FTLM
can remain correct even down to $T \to 0$. \cite{jaklic00}

\begin{figure}[h!]
\vspace{0.3cm}
\centering
\includegraphics[width=\columnwidth]{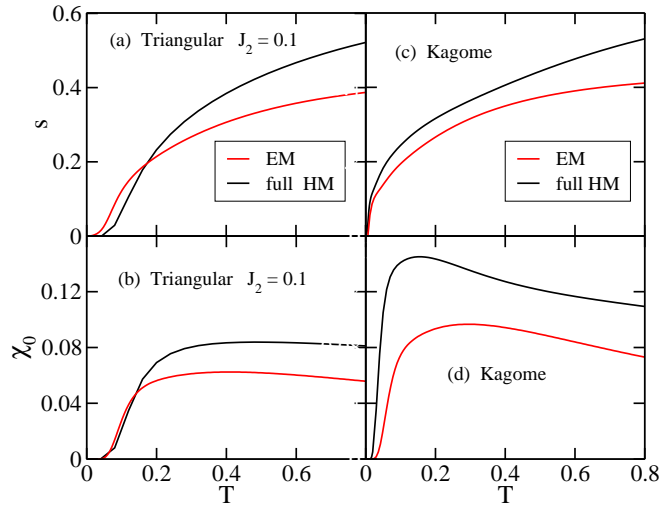}
\caption{ Results for the effective model (EM) for the triangular lattice, compared with the
full HM. All results for are for $N=30$ sites \cite{prelovsek18}: a)
for entropy density $s(T)$ and b) uniform susceptibility
$\chi_0(T)$.  c) and d) same quantities for the kagome lattice,
compared to full HM, all on $N=42$ sites \cite{schnack18}.}  \label{figsl1}
\end{figure}

\section{Entropy and uniform susceptibility} 

Let us first benchmark results within the EM with the existing results for
the full HM on TL and KL. In Figs.~\ref{figsl1}a,b we present $s(T)$ and
$\chi_0(T)$, respectively, as obtained on TL for $J_2=0.1$ on $N=30$
via FTLM on EM, compared with the full HM of the same size
\cite{prelovsek18}.  The qualitative behavior of both quantities
within EM is quite similar at low $T$ to the full HM, in particular for $s(T)$,
although EM evidently misses part of $s(T)$ with increasing $T$ due to reduced 
basis space. More pronounced is quantitative (but not qualitative) discrepancy 
in $\chi_0(T)$ which can be attributed to missing higher spin states in EM. 
Still, the peak in $\chi_0(T)$ and related spin (triplet) gap
$\Delta_t >0$ at low $T$) are reproduced well within EM.
Similar conclusions emerge from Figs.~\ref{figsl1}c,d
where corresponding results are compared for the KL, where full-HM
results for $s(T)$ and $\chi_0(T)$ are taken from study on $N=42$
sites \cite{schnack18}. Here, EM clearly reproduces reasonably not
only triplet gap $\Delta_t \sim 0.1$ but also singlet excitations
dominating $s(T \to 0)$, while apparently EM underestimates the value
of $\chi_0(T)$.

\begin{figure}[h!]
\vspace{0.4cm}
\centering
\includegraphics[width=0.9\columnwidth]{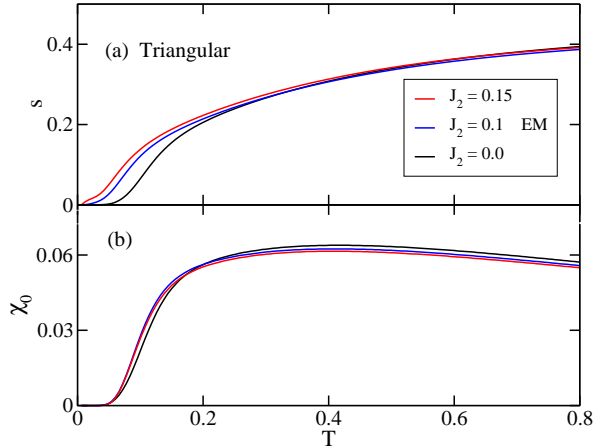}
\caption{ Results within EM on triangular lattice for different $J_2 =0, 0.1, 0.15$; a)
for $s(T)$, and b) $\chi_0(T)$.}
\label{figsl2}
\end{figure}

\begin{figure}[h!]
\vspace{0.4cm}
\centering
\includegraphics[width=0.9\columnwidth]{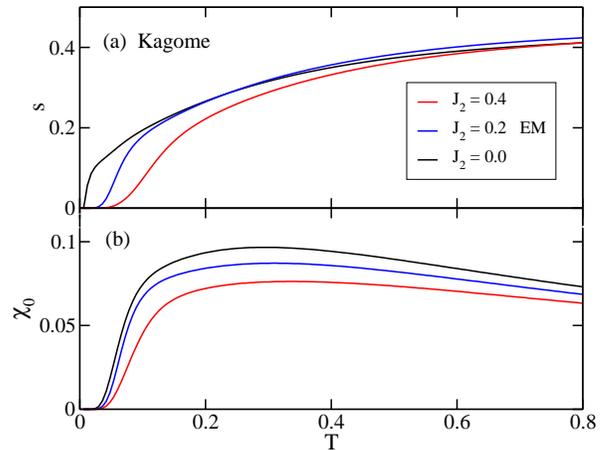}
\caption{ Results within EM on kagome  lattice 
for different $J_2 = 0, 0.2, 0.4$; a) for $s(T)$, and b) $\chi_0(T)$.}
\label{figsl3}
\end{figure}

After testing with the full model, we present in Figs.~\ref{figsl2}
and \ref{figsl3} EM
results for $s(T)$ and $\chi_0(T)$ for both lattices as they vary with
$J_2 > 0$.  In Fig.~\ref{figsl2}a,b we follow the behavior on TL for different
$J_2=0, 0.1, 0.15$.  From the inflection (vanishing second derivative)  point
of $s(T)$ defining singlet temperature $T=T_s$  one can speculate on the 
coherence scale  (in the case of LRO) or possible
(singlet) excitation gap $\Delta_s \lesssim T_s$ (in the case of SL),
at least provided that $T_s > T_{fs}$.  
Although the influence of $J_2 >0$ does not appear large, it still
introduces a qualitative difference.  From this perspective $s(T)$
within TL EM at $J_2=0$ reveals higher effective $T_s$ 
 being consistent with $s(T<T_s) \propto T^2$ and a spiral LRO at $T=0$. 
Still, we get in this case  $T_s \sim T_{fs}$ within
EM, so we can hardly make stronger conclusions.  

On the other hand, for TL and $J_2 = 0.1, 0.15$ where the SL can be expected
\cite{kaneko14,iqbal16,gong17} the EM reveals smaller $T_s \sim 0.05$
which is the signature of the singlet gap (which could still be
finite-size dependent).  More important, results confirm large
residual entropy $s(T) \sim 0.1 = 0.14 s_{max}$ even at $T \sim 0.1$.
This is in contrast with $\chi_0(T)$ in Fig.~\ref{figsl2}b which reveal
$T$-variation weakly dependent on $J_2$.  While for $J_2=0$ the drop
at $\chi_0(T < T_t)$ is the signature of the finite-size spin gap (where
due to magnetic LRO $\chi_0^0 = \chi_0( T \to 0) >0 $ is expected),
$J_2 =0.1, 0.15$ examples are different  since vanishing $\chi_0^0$
could indicate the spin triplet gap $\Delta_t > 0.1$ beyond the
finite-size effects, i.e. $T_t \sim 0.1 > T_{fs}$.  

In Figs.~\ref{figsl3}a,b we present the same quantities for the case of KL,
now for $J_2=0, 0.2, 0.4$.  The effect of $J_2>0$ is opposite, since
it is expected to recover the LRO at $J_2 \sim 0.4$ \cite{kolley15}, with
$120^0$ spin orientation analogous to $J_2=0$ TL. The largest low-$T$
entropy $s(T)$ is found for KL with $J_2=0$. Moreover, EM here yields
a quantitive agreement with the full HM \cite{schnack18}, revealing
large remanent $s(T)$ due to singlet (chirality) excitations down to
$T \sim 0.01$ \cite{lauchli19}.  The evident
effect of $J_2 >0$ is to reduce $s(T)$ and finally leading
$s(T) \propto T^2$ at large $J_2 \sim 0.4$ which should be a regime of
magnetic LRO \cite{kolley15}. Again, at $J_2=0$ in contrast to entropy
$\chi_0(T)$ has well pronounced downturn at
$T \sim 0.1$ consistent with the triplet gap $\Delta_t \sim 0.1$ found
in most other numerical studies \cite{sindzingre00,misguich07,
  bernu15,schnack18,lauchli19}.  Introducing $J_2 > 0$ does not change
$\chi_0(T)$ qualitatively.

\section{Wilson ratio: results}  

\begin{figure}[h!]
\vspace{0.3cm}
\centering
\includegraphics[width=0.7\columnwidth]{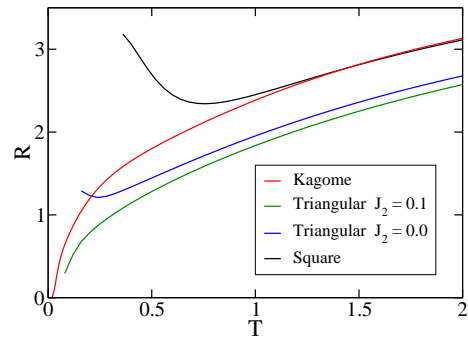}
\caption{ Wilson ratio $R(T)$, evaluated from $s(T)$ and $\chi_0(T)$ for  full HM 
on square lattice, triangular lattice with $J_2=0, 0.1$ \cite{prelovsek18}, and kagome 
lattice \cite{schnack18}. Results are presented for $T>T_{fs}$.}
\label{figsl4}
\end{figure}

To calculate $R(T)$, Eq.~(\ref{rw}), let us first use available
results for the full HM for TL
\cite{prelovsek18} and KL \cite{schnack18}, comparing in Fig.~\ref{figsl4} also
the result for unfrustrated HM on a square lattice \cite{prelovsek18}.
Here, we take into account data for $T> T_{fs}$, acknowledging that
$T_{fs}$ are quite different (taking $s(T_{fs}) \sim 0.1$ as
criterion) for these systems, representative also for the degree of
frustration. Fig.~\ref{figsl4} already confirms different scenarios for
$R(T)$. In  HM on a simple square lattice, starting from high-$T$ limit $R(T)$
reaches minimum at $T^* \sim 0.7$ and then increases, which is consistent with
$R(T\to 0) \to \infty$ for a 2D system with $T=0$ magnetic LRO. The same
behavior appears for TL at $J_2=0$ with a shallow minimum shifted to
$T^* \sim 0.3$. In contrast, results for KL as well as for TL with
$J_2=0.1$ do not reveal such increase, at least not for $T>T_{fs}$,
and they are more consistent with the interpretation that
$R(T\to 0) \to 0$.

\begin{figure}[h!]
\vspace{0.4cm}
\centering
\includegraphics[width=0.9\columnwidth]{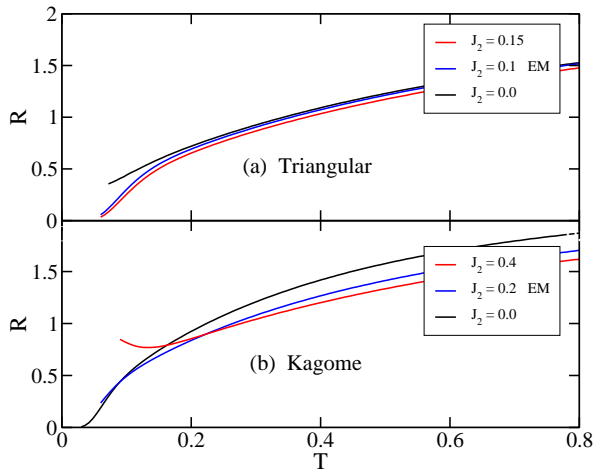}
\caption{ $R(T)$, evaluated within EM for: a) TL with $J_2=0,0.1,0.15$,
and b) KL for $J_2 = 0, 0.2, 0.4$.}
\label{figsl5}
\end{figure}

Finally, results for $R(T)$ within EM are shown in Fig.~\ref{figsl5}a,b as
they follow from Figs.~\ref{figsl2} and \ref{figsl3} for different $J_2 \geq 0$. We recognize
that EM qualitatively reproduce numerical data within the full HM on
Fig.~\ref{figsl4}. Although for $J_2=0$ TL results in Fig.~\ref{figsl5}a fail to reveal
clearly the minimum down to $T_{fs}\sim 0.1$, there is still a marked
difference to the SL regime $J_2=0.1, 0.15$ where EM confirms
$R_0 \ll 1$. Results within EM for KL, as shown in
Fig.~\ref{figsl5}b, are even better demonstration for vanishing $R_0$. Here, for $J_2=0$ EM
yields quite similar $R(T)$, decreasing and tending towards
$R_0 \sim 0$.  On the other hand, the effect of finite $J_2 >0$ is
well visible and leads towards magnetic LRO with $R_0 \to \infty$ for
$J_2 =0.4$.
   
\section{Conclusions} 

The main message (apparently not stressed in previous published studies of the SL
models) of presented results for 
the thermodynamic quantities: the entropy $s(T)$,
susceptibility $\chi_0(T)$, and in particular the Wilson ratio $R(T)$,
behave similarity in the extended $J_1$-$J_2$ Heisenberg model 
on TL and KL in their presumed SL regimes. Moreover, results on both lattices
follow quite analogous development by varying the nnn exchange $J_2 >0$, whereby
the effect is evidently opposite between TL and KL, regarding the magnetic LRO and SL 
phases \cite{kolley15}. 

While above similarities can be extracted already from full-model results obtained via 
FTLM \cite{schnack18,prelovsek18}, the introduction of reduced effective model appears 
crucial  for the understanding and useful analytical insight.  Apart from offering some
numerical advantages of reduced Hilbert space of basis states, essential for
ED methods, EM clearly puts into display two (separate) degrees of freedom: a) 
the  effective spin $s=1/2$ degrees ${\bf s}_i$,
determining $\chi_0(T)$, but as well the dynamical quantities, as e.g. the dynamical spin structure factor
$S({\bf q}, \omega)$ not discussed here \cite{prelovsek18}, 
b) chirality pseudospin degrees ${\bf \tau}_i$ which do
not contribute to $\chi_0(T)$, but enter the entropy $s(T)$ and related
specific heat. From the EM and its dependence on $J_2$ it
is also quite evident where to expect large fluctuations of  ${\bf \tau}_i$
and consequently SL regime, which is not very apparent within the full HM.
As the EM is based on the direct reduction/projection  of the basis (and not on perturbation
expansion) of the original HM, it is expected that the correspondence is qualitative and not fully
quantitative. In any case, the EM is also by itself a valuable and highly nontrivial model,
and could serve as such to understand better the onset and properties of SL. 

The essential common feature of SL regimes in HM on
both lattices is a pronounced remanent $s(T)>0$ at $T \ll J_1$, which
within the EM has the origin in dominant low-energy chiral
fluctuations, well below the effective spin triplet gap $\Delta_t$
which is revealed by the drop of $\chi_0(T)$.  As a consequence we
observe the vanishing of the Wilson ratio $R_0 = R(T\to 0) \to 0$,
which seems to be quite generic feature of 2D SL models \cite{prelovsek19}. 
Clearly, due to finite-size
restrictions we could hardly distinguish a spin-gapped system from
scenarios with more delicate gap structure which could also lead to
renormalized $R_0 \ll 1$.  Moreover, it is even harder to decide
beyond finite-size effects whether singlets excitations are gapless or with finite 
singlet gap $\Delta_s >0$, which  should be in any case very small $\Delta_s \ll J$ and from
results  $R(T\to 0) \to 0$ have to be evidently smaller than a triplet one, 
i.e. $\Delta_s<\Delta_t$.  From our finite-size results it is also hard to exclude the 
scenario of valence-bond ordered ground state (with broken translational symmetry),
although we do not see an indication for that and vanishing $R_0$, it is also 
not easily compatible with the latter either.

Quantities discussed above are measurable in real
materials and have been indeed discussed for some of them. There are
evident experimental difficulties, i.e., $\chi_0(T)$ can have
significant impurity contributions while $s(T)$ may be masked by
phonon contribution at $T > 0$.  The essential hallmark for material
candidates for the presented SL scenario should be a substantial
entropy $s(T)$ persisting well below $T \ll J_1$.  There are indeed
several studies of $s(T)$ reported for different SL candidates (with
some of them revealing transitions to magnetic LRO at very low $T$), e.g., for
KL systems volborthite \cite{hiroi01}, YCu$_3$(OH)$_6$Cl$_3$
\cite{zorko19,arh19}, and recent TL systems 1T-TaS$_2$
\cite{kratochvilova17} and Co-based SL materials \cite{zhong19}.
While existing experimental results on SL materials do not seem to indicate
vanishing (or very small)  $R_0$, it might also happen that
the (above) considered SL models are not fully capturing  the low-$T$ physics. 
In particular, there could be important role played by additional terms, e.g., the 
Dzaloshinski-Moriya interaction \cite{cepas08,rousochatzakis09,arh19} and/or 3D 
coupling, which can reduce $s(T)$ or even induce magnetic LRO at $T \to 0$.

\acknowledgments This work is supported by the program P1-0044 of the
Slovenian Research Agency. Authors thank J. Schnack for providing
their data for kagome lattice and  J. Schnack, F. Becca,
A. Zorko, K. Morita and T. Tohyama for fruitful discussions.

\bibliography{manuslprb}

\end{document}